

\rightline{UM-TH 94--01}\par\noindent
\rightline{hep-ph/9401275}\par\noindent

\def\pmb#1{\setbox0=\hbox{#1}%
  \hbox{\kern-.025em\copy0\kern-\wd0
  \kern.05em\copy0\kern-\wd0
  \kern-0.025em\raise.0433em\box0} }

\catcode`@=11
\def\leftrightarrowfill{$\m@th\mathord\leftarrow \mkern-6mu
  \cleaders\hbox{$\mkern-2mu \mathord- \mkern-2mu$}\hfill
  \mkern-6mu \mathord\rightarrow$}
\def\overleftrightarrow#1{\vbox{\ialign{##\crcr
     \leftrightarrowfill\crcr\noalign{\kern-1pt\nointerlineskip}
     $\hfil\displaystyle{#1}\hfil$\crcr}}}
\catcode`@=12

\def\approxge{\hbox {\hfil\raise .4ex\hbox{$>$}\kern-.75 em
\lower .7ex\hbox{$\sim$}\hfil}}
\def\approxle{\hbox {\hfil\raise .4ex\hbox{$<$}\kern-.75 em
\lower .7ex\hbox{$\sim$}\hfil}}

\def \abstract#1 {\vskip 0.5truecm\sepline\vskip 0.5truecm
$$\vbox{\hsize=15truecm\noindent #1}$$}
\def \SISSA#1#2 {\vfil\vfil\centerline{Ref. S.I.S.S.A. #1 CM (#2)}}
\def \PACS#1 {\vfil\line{\hfil\hbox to 15truecm{PACS numbers: #1 \hfil}\hfil}}

\def \hfigure
     #1#2#3       {\midinsert \vskip #2 truecm $$\vbox{\hsize=14.5truecm
             \seven\baselineskip=10pt\noindent {\bcp \noindent Figure  #1}.
                   #3 } $$ \vskip -20pt \endinsert }

\def \hfiglin
     #1#2#3       {\midinsert \vskip #2 truecm $$\vbox{\hsize=14.5truecm
              \seven\baselineskip=10pt\noindent {\bcp \hfil\noindent
                   Figure  #1}. #3 \hfil} $$ \vskip -20pt \endinsert }

\def \vfigure
     #1#2#3#4     {\dimen0=\hsize \advance\dimen0 by -#3truecm
                   \midinsert \vbox to #2truecm{ \seven
                   \parshape=1 #3truecm \dimen0 \baselineskip=10pt \vfill
                   \noindent{\bcp Figure #1} \pretolerance=6500#4 \vfill }
                   \endinsert }

%
\def \ref
     #1#2         {\smallskip \item{[#1]}#2}
\def \sepline     {\medskip\centerline{\vbox{\hrule width5truecm}} \medskip}

\def \tabrul2     {\noalign{\vskip 5truept \hrule \vskip 2truept \hrule
                   \vskip 5truept} }


\footline={\ifnum\pageno>0 \tenrm \hss \folio \hss \fi }

\def\today
 {\count10=\year\advance\count10 by -1900 \number\day--\ifcase
  \month \or Jan\or Feb\or Mar\or Apr\or May\or Jun\or
             Jul\or Aug\or Sep\or Oct\or Nov\or Dec\fi--\number\count10}

\def\hour{\count10=\time\count11=\count10
\divide\count10 by 60 \count12=\count10
\multiply\count12 by 60 \advance\count11 by -\count12\count12=0
\number\count10 :\ifnum\count11 < 10 \number\count12\fi\number\count11}

\def\draft{
   \baselineskip=20pt
   \def\makeheadline{\vbox to 10pt{\vskip-22.5pt
   \line{\vbox to 8.5pt{}\the\headline}\vss}\nointerlineskip}
   \headline={\hfill \seven {\bcp Draft version}: today is \today\ at \hour
              \hfill}
          }

%
%

%
\catcode`@=11
%
%
\def\b@lank{ }

\newif\if@simboli
\newif\if@riferimenti

\newwrite\file@simboli
\def\simboli{
    \immediate\write16{ !!! Genera il file \jobname.SMB }
    \@simbolitrue\immediate\openout\file@simboli=\jobname.smb}

\newwrite\file@ausiliario
\def\riferimentifuturi{
    \immediate\write16{ !!! Genera il file \jobname.AUX }
    \@riferimentitrue\openin1 \jobname.aux
    \ifeof1\relax\else\closein1\relax\input\jobname.aux\fi
    \immediate\openout\file@ausiliario=\jobname.aux}

\newcount\eq@num\global\eq@num=0
\newcount\sect@num\global\sect@num=0

\newif\if@ndoppia
\def\numerazionedoppia{\@ndoppiatrue\gdef\la@sezionecorrente{\the\sect@num}}

\def\se@indefinito#1{\expandafter\ifx\csname#1\endcsname\relax}
\def\spo@glia#1>{} 

\newif\if@primasezione
\@primasezionetrue

\def\s@ection#1\par{\immediate
    \write16{#1}\if@primasezione\global\@primasezionefalse\else\goodbreak
    \vskip\spaziosoprasez\fi\noindent
    {\bf#1}\nobreak\vskip\spaziosottosez\nobreak\noindent}
%

\def\sezpreset#1{\global\sect@num=#1
    \immediate\write16{ !!! sez-preset = #1 }   }

\def\spaziosoprasez{50pt plus 60pt}
\def\spaziosottosez{15pt}

\def\sref#1{\se@indefinito{@s@#1}\immediate\write16{ ??? \string\sref{#1}
    non definita !!!}
    \expandafter\xdef\csname @s@#1\endcsname{??}\fi\csname @s@#1\endcsname}


\def\adv#1{\global\advance\sect@num by #1
           \global\eq@num=0}


\def\autosez#1#2\par{
    \global\advance\sect@num by 1\if@ndoppia\global\eq@num=0\fi
    \xdef\la@sezionecorrente{\the\sect@num}
    \def\usa@getta{1}\se@indefinito{@s@#1}\def\usa@getta{2}\fi
    \expandafter\ifx\csname @s@#1\endcsname\la@sezionecorrente\def
    \usa@getta{2}\fi
    \ifodd\usa@getta\immediate\write16
      { ??? possibili riferimenti errati a \string\sref{#1} !!!}\fi
    \expandafter\xdef\csname @s@#1\endcsname{\la@sezionecorrente}
    \immediate\write16{\la@sezionecorrente. #2}
    \if@simboli
      \immediate\write\file@simboli{ }\immediate\write\file@simboli{ }
      \immediate\write\file@simboli{  Sezione
                                  \la@sezionecorrente :   sref.   #1}
      \immediate\write\file@simboli{ } \fi
    \if@riferimenti
      \immediate\write\file@ausiliario{\string\expandafter\string\edef
      \string\csname\b@lank @s@#1\string\endcsname{\la@sezionecorrente}}\fi
    \goodbreak\vskip 48pt plus 60pt
\centerline{\lltitle #2}                     
\par\nobreak\vskip 15pt \nobreak\noindent}

\def\semiautosez#1#2\par{
    \gdef\la@sezionecorrente{#1}\if@ndoppia\global\eq@num=0\fi
    \if@simboli
      \immediate\write\file@simboli{ }\immediate\write\file@simboli{ }
      \immediate\write\file@simboli{  Sezione ** : sref.
          \expandafter\spo@glia\meaning\la@sezionecorrente}
      \immediate\write\file@simboli{ }\fi
\noindent\lltitle \s@ection#2 \par}


\def\eqpreset#1{\global\eq@num=#1
     \immediate\write16{ !!! eq-preset = #1 }     }

\def\eqref#1{\se@indefinito{@eq@#1}
    \immediate\write16{ ??? \string\eqref{#1} non definita !!!}
    \expandafter\xdef\csname @eq@#1\endcsname{??}
    \fi\csname @eq@#1\endcsname}

\def\eqlabel#1{\global\advance\eq@num by 1
    \if@ndoppia\xdef\il@numero{\la@sezionecorrente.\the\eq@num}
       \else\xdef\il@numero{\the\eq@num}\fi
    \def\usa@getta{1}\se@indefinito{@eq@#1}\def\usa@getta{2}\fi
    \expandafter\ifx\csname @eq@#1\endcsname\il@numero\def\usa@getta{2}\fi
    \ifodd\usa@getta\immediate\write16
       { ??? possibili riferimenti errati a \string\eqref{#1} !!!}\fi
    \expandafter\xdef\csname @eq@#1\endcsname{\il@numero}
    \if@ndoppia
       \def\usa@getta{\expandafter\spo@glia\meaning
       \la@sezionecorrente.\the\eq@num}
       \else\def\usa@getta{\the\eq@num}\fi
    \if@simboli
       \immediate\write\file@simboli{  Equazione
            \usa@getta :  eqref.   #1}\fi
    \if@riferimenti
       \immediate\write\file@ausiliario{\string\expandafter\string\edef
       \string\csname\b@lank @eq@#1\string\endcsname{\usa@getta}}\fi}

\def\autoeqno#1{\eqlabel{#1}\eqno(\csname @eq@#1\endcsname)}
\def\autoleqno#1{\eqlabel{#1}\leqno(\csname @eq@#1\endcsname)}
\def\eqrefp#1{(\eqref{#1})}


\def\eq{\autoeqno}
\def\req{\eqrefp}



\newcount\cit@num\global\cit@num=0

\newwrite\file@bibliografia
\newif\if@bibliografia
\@bibliografiafalse

\def\lp@cite{[}
\def\rp@cite{]}
\def\trap@cite#1{\lp@cite #1\rp@cite}
\def\lp@bibl{[}
\def\rp@bibl{]}
\def\trap@bibl#1{\lp@bibl #1\rp@bibl}

\def\refe@renza#1{\if@bibliografia\immediate        
    \write\file@bibliografia{
    \string\item{\trap@bibl{\cref{#1}}}\string
    \bibl@ref{#1}\string\bibl@skip}\fi}

\def\ref@ridefinita#1{\if@bibliografia\immediate\write\file@bibliografia{
    \string\item{?? \trap@bibl{\cref{#1}}} ??? tentativo di ridefinire la
      citazione #1 !!! \string\bibl@skip}\fi}

\def\bibl@ref#1{\se@indefinito{@ref@#1}\immediate
    \write16{ ??? biblitem #1 indefinito !!!}\expandafter\xdef
    \csname @ref@#1\endcsname{ ??}\fi\csname @ref@#1\endcsname}

\def\c@label#1{\global\advance\cit@num by 1\xdef            
   \la@citazione{\the\cit@num}\expandafter
   \xdef\csname @c@#1\endcsname{\la@citazione}}

\def\bibl@skip{\vskip +4truept}


\def\stileincite#1#2{\global\def\lp@cite{#1}\global   
    \def\rp@cite{#2}}                                 
\def\stileinbibl#1#2{\global\def\lp@bibl{#1}\global   
    \def\rp@bibl{#2}}                                 

\def\citpreset#1{\global\cit@num=#1
    \immediate\write16{ !!! cit-preset = #1 }    }

\def\autobibliografia{\global\@bibliografiatrue\immediate
    \write16{ !!! Genera il file \jobname.BIB}\immediate
    \openout\file@bibliografia=\jobname.bib}

\def\cref#1{\se@indefinito                  
   {@c@#1}\c@label{#1}\refe@renza{#1}\fi\csname @c@#1\endcsname}

\def\cite#1{\trap@cite{\cref{#1}}}                  
\def\ccite#1#2{\trap@cite{\cref{#1},\cref{#2}}}     
\def\ncite#1#2{\trap@cite{\cref{#1}--\cref{#2}}}    
\def\upcite#1{$^{\,\trap@cite{\cref{#1}}}$}               
\def\upccite#1#2{$^{\,\trap@cite{\cref{#1},\cref{#2}}}$}  
\def\upncite#1#2{$^{\,\trap@cite{\cref{#1}-\cref{#2}}}$}  

\def\clabel#1{\se@indefinito{@c@#1}\c@label           
    {#1}\refe@renza{#1}\else\c@label{#1}\ref@ridefinita{#1}\fi}

\def\biblskip#1{\def\bibl@skip{\vskip #1}}           

\def\insertbibliografia{\if@bibliografia             
    \immediate\write\file@bibliografia{ }
    \immediate\closeout\file@bibliografia
    \catcode`@=11\input\jobname.bib\catcode`@=12\fi}


\def\commento#1{\relax}
\def\biblitem#1#2\par{\expandafter\xdef\csname @ref@#1\endcsname{#2}}


\catcode`@=12



\tolerance 100000
\biblskip{+8truept}                        
\def\hbup{\hfill\break\baselineskip 14pt}  


\global\newcount\notenumber \global\notenumber=0
\def\note #1 {\global\advance\notenumber by1 \baselineskip 10pt
              \footnote{$^{\the\notenumber}$}{\nine #1} \interlinea}






\def\gtrsim{\ \rlap{\raise 2pt \hbox{$>$}}{\lower 2pt \hbox{$\sim$}}\ }
\def\lesssim{\ \rlap{\raise 2pt \hbox{$<$}}{\lower 2pt \hbox{$\sim$}}\ }


\def\mn{\medskip\noindent}

\def\o{\over}



\def\ea{{\elevenit et.al.}}
\def\ib{{\elevenit ibid.\ }}

\def\npb#1{{\elevenit Nucl. Phys.} {\elevenbf B#1},}
\def\plb#1{{\elevenit Phys. Lett.} {\elevenbf B#1},}
\def\prd#1{{\elevenit Phys. Rev.} {\elevenbf D#1},}
\def\prl#1{{\elevenit Phys. Rev. Lett.} {\elevenbf #1},}
\def\ncim#1{{\elevenit Nuo. Cim.} {\elevenbf #1},}

\def\prep#1{{\elevenit Phys. Rep.} {\elevenbf #1},}

\def\ijmpa#1{{\elevenit Int. Jour. Mod. Phys.} {\elevenbf A#1},}


\stileincite{}{}     
\stileinbibl{}{.}    

\numerazionedoppia   



\def\o{\over}


\def\G{{\cal G_{\rm SM}}}
\def\E{{\rm E}_6}


\def\pr{\prime}



\def\nue{{$\nu_e\,$}}
\def\num{{$\nu_\mu\,$}}
\def\nut{{$\nu_\tau\,$}}

\def\E#1{{$E_{#1}\,$}}













\def\lu{{H^cQu^c}}
\def\ld{{HQd^c}}
\def\lt{{HLe^c}}
\def\lq{{S^chh^c}}
\def\lc{{hu^ce^c}}
\def\ls{{LQh^c}}
\def\lst{{\nu^chd^c}}
\def\lo{{hQQ}}
\def\ln{{h^cu^cd^c}}
\def\ldi{{H^cL\nu^c}}
\def\luu{{H^cHS^c}}


\def\l#1#2#3#4{\lambda^{^{\scriptstyle (#1)}}_{\scriptscriptstyle #2#3#4}\>}
\def\lw#1{\lambda^{^{\scriptstyle (#1)}}\>}


\def\ie{{\it i.e.\ }}
\def\eg{{\it e.g.\ }}

\def\ra{\rangle}
\def\la{\langle}

\def\o{\over}


\def\nue{{$\nu_e\,$}}
\def\num{{$\nu_\mu\,$}}
\def\nut{{$\nu_\tau\,$}}

\def\E#1{{$E_{#1}\,$}}


\def\lu{{H^cQu^c}}
\def\ld{{HQd^c}}
\def\lt{{HLe^c}}
\def\lq{{S^chh^c}}
\def\lc{{hu^ce^c}}
\def\ls{{LQh^c}}
\def\lst{{\nu^chd^c}}
\def\lo{{hQQ}}
\def\ln{{h^cu^cd^c}}
\def\ldi{{H^cL\nu^c}}
\def\luu{{H^cHS^c}}


\def\l#1#2#3#4{\lambda^{^{\scriptstyle (#1)}}_{\scriptscriptstyle #2#3#4}\>}
\def\lw#1{\lambda^{^{\scriptstyle (#1)}}\>}

\def\G{{\cal G_{\rm SM}}}
\def\E{{\rm E}_6}

\def\pr{\prime}




%
\headline={\ifnum\pageno=1\firstheadline\else
\ifodd\pageno\rightheadline \else\leftheadline\fi\fi}
\def\firstheadline{\hfil}
\def\rightheadline{\hfil}
\def\leftheadline{\hfil}
        \footline={\ifnum\pageno=1\firstfootline\else\otherfootline\fi}
\def\firstfootline{\rm\hss\folio\hss}
\def\otherfootline{\hfil}

\font\tenrm=cmr10
\font\tenit=cmti10
\font\elevenbf=cmbx10 scaled\magstep 1
\font\elevenrm=cmr10 scaled\magstep 1
\font\elevenit=cmti10 scaled\magstep 1


\autobibliografia
\nopagenumbers
\line{\hfil }
\vglue 1cm
\hsize=6.0truein
\vsize=8.5truein
\parindent=3pc
\baselineskip=10pt
\centerline{\elevenbf NEUTRINO PHENOMENOLOGY FROM}
\vglue 0.2cm
\centerline{\elevenbf UNCONVENTIONAL E$_{\bf 6}$
MODELS$^{\displaystyle\dagger}$ }
\vglue 0.2cm
\vglue 1.0cm
\centerline{\tenrm ENRICO NARDI }
\baselineskip=13pt
\centerline{\tenit Randall Laboratory of Physics, University of Michigan}
\baselineskip=12pt
\centerline{\tenit Ann Arbor, MI 48109-1120, U.S.A.}

\vglue 2.5cm
\centerline{\tenrm ABSTRACT}
\vglue 0.3cm
{\rightskip=3pc
 \leftskip=3pc
\elevenrm\baselineskip=14pt
 \noindent
Superstring derived $\E$ models can accommodate small neutrino masses
if  a discrete symmetry is imposed which forbids tree level Dirac
neutrino masses but allows for radiative mass generation. The only
possible symmetries of this kind are known to be generation dependent.
Thus we explore the possibility that,  as a consequence of such a
symmetry, the three sets of light states in each generation do not
have the same assignments with respect to the {\bf 27} of $E_6$,
implying that the gauge interactions under the additional $U(1)^\pr$
factors are non universal. Models realising such a scenario are
viable, and by requiring the number of light neutral states to be
minimal, an almost unique pattern of neutrino masses and mixings
arises. We briefly discuss a model in which, with a natural choice of
the parameters, $m_{\nu_\tau}\sim 0.1-10\,$eV is generated at one
loop, $m_{\nu_\mu}\sim 10^{-3}\,$eV  is generated at two loops and
${\nu_e}$ remains massless.}

\vfill\noindent
--------------------------------------------\phantom{-} \hfil\break
\medskip
\leftline{$^\dagger$ Talk given at the
{\elevenit ``International School on Cosmological Dark Matter''},}
\leftline{\phantom{$^\dagger$} Valencia, Spain, October 4-8, 1993.}
\vglue 1truecm
\leftline{E-mail: nardi@umiphys.bitnet}
\bigskip
\leftline{UM-TH 94--01}
\bigskip
\centerline{January 1994}
\eject
\null

\vglue 0.8cm
\adv{1}
\line{\elevenbf 1. Introduction \hfil}
\bigskip
\def\interlinea{\baselineskip=14pt}
\baselineskip=14pt
\elevenrm
\noindent
It is generally believed that neutrinos possess very small but
non-vanishing masses.
While there is no fundamental reason for the neutrinos to be exactly
massless, small $\nu$ masses are needed in any particle physics
explanation of the solar neutrino problem, and at the same time they
imply several interesting phenomenological consequences.
A  very attractive way of generating naturally small neutrino masses
is through the use of the see-saw mechanism\upcite{see-saw}. In $\E$
supersymmetric Grand Unified Theories (GUTs)\upcite{rizzo-e6}, as
derived from superstring theories, the see-saw mechanism cannot be
easily implemented since the Higgs representation necessary to
generate a large Majorana mass for the right-handed neutrinos is
absent.
However, even in the absence of Majorana terms, small masses can be
generated through radiative corrections in models in which at the
lowest order $m_\nu=0$. As was pointed out by Campbell et.
al.\upcite{ellis-e6} and Masiero et al.\upcite{MNS}, $\E$ GUTs do offer
the possibility of implementing this second mechanism.

The fermion content of models based on $\E$ is enlarged with respect
to the Standard Model (SM). In fact two additional lepton
$SU(2)$-doublets, two $SU(2)$-singlet neutral states and two
color-triplet $SU(2)$-singlet $d$-type quarks are present in
the fundamental representation of the group.
In order to forbid neutrino masses at the tree level an appropriate
discrete symmetry has to be imposed on the superpotential of the
model.  Branco and
Geng \upcite{branco} have shown that no generation-blind symmetry
exists that forbids non vanishing neutrino masses at the tree level,
and at the same time allows for the radiative generation of the masses
at one loop. As a result, in order to implement this mechanism a
symmetry that does not act in the same way on the three generations is
needed.

It was recently pointed out\upcite{f}
that once we chose
to build a model based on a symmetry that does distinguish among the
different generations, there is no reason in principle to expect that
this symmetry will result in a set of
{\elevenit light} fermions  ({\ie the known states) that will
exactly replicate throughout the three generations.
To state this idea more clearly, we wish to suggest the possibility
that what we call ``\nut"
is actually assigned to an  $SU(2)$ doublet which has a different
embedding in $\E$ with respect to the doublet that contains
what we call ``\nue".
In the following we will denote this kind of
non-standard embeddings as `unconventional
assignments' (UA).
As a consequence the two neutrinos
will have different $\E$ gauge interactions.
Obviously, experimentally  we know that the $SU(2)\times U(1)$
interactions of the fermions do respect universality
with a high degree of precision,
however, in the class of models that we want to investigate one or two
additional $U(1)^\pr$ abelian factors are always present, implying
additional massive neutral gauge bosons possibly at energies $O$(TeV)
or less. The possibility that the $U(1)^\pr$ interactions of the
known fermions could violate universality then is indeed still
phenomenologically viable.
In Section 2 we will briefly
outline a scenario that realises this idea. A more complete
description of the theoretical framework can be found in
Ref. [\cite{f}].
In Section 3  we will concentrate on the neutrino phenomenology,
and we will describe the pattern of masses and mixings that
is predicted by our scheme, and in Section 4 we will draw the conclusions.
We believe that the unconventional scenario that we are going to
analyse here could be interesting in itself, since it is not a priori
obvious that models in which the `low' energy gauge interactions of the known
fermions are not universal can be consistently constructed. However, it
turns out that beyond being viable, these models also lead to an
interesting phenomenology, expecially in the neutrino sector, and
as well imply some rather unusual consequences.
For example a few peculiar effects in
the propagation of the neutrinos through matter could arise,
and have been discussed in Ref. [\cite{f}].
Since neutrinos come in doublets with L-handed charged leptons,
UA for the
$\nu$'s also imply that the neutral current interactions for
$e_L$, $\mu_L$ and $\tau_L$ can be different.
This will result in deviations from unity for the rate of production
of different lepton flavors, and represents the most clean signature
of the UA models.
Such a signature could be most easily detected
in $e^+ e^-$ annihilation at high c.m. energy, as for example at
LEP II and at a 500 GeV Next Linear Collider\upcite{fphen}.
However, in order to identify completely the correct pattern of UA,
the measurement of a large set of quantities is needed. A thorough
analysis of the effect of UA on various cross sections and
asymmetries, aiming to study the possibility of identifying
unequivocally the different possible embeddings, can be found in
Ref. [\cite{e6asy}].

\vglue 0.6cm
\adv{1}
\line{\elevenbf 2. Unconventional Assignments in E$_{\bf 6}$ models. \hfil}
\vglue 0.4cm
\noindent

In $\E$ grand unified theories,
as many as two new neutral gauge bosons
can be present, corresponding to the two additional
Cartan generators that are not present in the SM gauge group.
Here we will consider the embedding of the SM gauge group
$$
\G\equiv SU(3)_c\times SU(2)_L\times U(1)_Y  \eq{2.1}
$$
in $\E$
through the maximal subalgebras chain:

\setbox3=\hbox{\phantom{+}\raise6pt\hbox{\big\vert\kern-.7mm\lower6pt\hbox
{$\longrightarrow \ U(1)_{\chi} \times SU(5)$}}}
\setbox4=\hbox{\phantom{+}\raise6pt\hbox{\big\vert\kern-.7mm\lower6pt\hbox
{$\longrightarrow \ \G $.}}}

$$
\eqalign{\E \ \longrightarrow \ U(1)_{\psi} \times SO&(10)  \cr
&\copy3 \cr
&\phantom{\longrightarrow \ U(1)_{\chi} \times SU(}
\copy4 \cr}                                  \eq{2.2}
$$
\noindent
In general the two additional neutral gauge boson
will correspond to some linear combinations
of the $U_\chi(1)$ and $U_\psi(1)$ generators that we will parametrize
in term of an angle $\theta$ as
$$
\eqalign{
&Z_\theta^\pr= Z_\psi \cos\theta - Z_\chi \sin\theta  \cr
&Z_\theta^{\pr\pr}= Z_\psi \sin\theta + Z_\chi \cos\theta. \cr}
\eq{2.3}
$$
\medskip\noindent
The angle $\theta$ is a model dependent parameter
whose value is determined by
the details of the breaking of the gauge symmetry.
In the following we will denote the
lightest of the two new gauge bosons as $Z_\theta$.

In the kind of models we are considering here, each generation of
matter fields belong to one fundamental {\bf 27}
representation of the group. The {\bf 27} branches to
the ${\bf 1} + {\bf 10}
+ {\bf 16}$ representations of $SO(10)$.
The known particles of the three generations,
together with an $SU(2)$ singlet neutrino ``$\nu^c$",  are usually
assigned to the {\bf 16} of $SO(10)$, that in turn branches to
${\bf 1}_{\bf 16}$ + ${\bf \bar 5}_{\bf 16}$  + ${\bf 10}_{\bf 16}$
of $SU(5)$.
Giving in parenthesis the Abelian charges
$Q_\psi$ and $Q_\chi$ for the different $SU(5)$ multiplets,
we have
\smallskip
$$
\eqalign{
&[{\bf 1}_{\bf 16}]\ (1c_\psi)\>(-5c_\chi)=
\big[\nu^c\big]                                          \cr
&[{\bf \bar 5}_{\bf 16}]\ (1c_\psi)\>(3c_\chi)=
\big[L={\nu \choose e}, \, d^c \big]                        \cr
&[{\bf 10}_{\bf 16}]\ (1c_\psi)\>(-1c_\chi)=
\big[Q = {u\choose d}, \, u^c, \, e^c \big]        \cr}
\eq{2.4}
$$
\medskip\noindent
The {\bf 10} of $SO(10)$ that branches to {\bf 5}$_{\bf 10}$ +
${\bf \bar 5}_{\bf 10}$  of $SU(5)$ contains the fields
$$
\eqalign{
&[{\bf 5}_{\bf 10}]\ (-2c_\psi)\>(-2c_\chi)=
\big[H = {{N\choose E}}, \,  h^c\big]     \cr
&[{\bf \bar 5}_{\bf 10}]\ (-2c_\psi)\>(2c_\chi)=
\big[H^c = {{E^c\choose N^c}}, \,h].       \cr}
\eq{2.5}
$$
Finally the singlet {\bf 1} of $SO(10)$ corresponds to
$$
[\>{\bf 1}_{\bf 1}\>]\ (4c_\psi)\>(0c_\chi) = [S^c].
\eq{2.6}
$$
According to the normalization
$\sum_{f=1}^{27}(Q_{\psi,\chi}^f)^2 =
\sum_{f=1}^{27}({1 \over 2}Y^f)^2=5$,
in \req{2.4}-\req{2.6}
we have respectively
$c_\psi = {1\o 6}\sqrt{5\o 2}$ and $c_\chi =
{1\o 6}\sqrt{2\o 3}$.
Matter fields will couple for example to the
$Z_\theta^\pr$ boson through the charge
$$
Q_\theta = Q_\psi \cos\theta - Q_\chi \sin\theta.  \eq{2.7}
$$
\noindent
The most general renormalizable superpotential
arising from the coupling of
the three {\bf 27}'s in \req{2.4}-\req{2.6}
and invariant under the low energy
gauge group \req{2.2} is\upcite{e6-super}
$$
W=W_1+W_2+W_3+W_4
$$
\noindent where
$$
\eqalign{
&W_1=\lw1\lu+\lw2\ld+\lw3\lt+\lw4\lq \cr
&W_2=\lw5\lc+\lw6\ls+\lw7\lst        \cr
&W_3=\lw8\lo+\lw9\ln                 \cr
&W_4=\lw{10}\ldi+\lw{11}\luu.          \cr
}
\eq{2.8}
$$
\mn
The Yukawa couplings in \req{2.8} are three index tensors in
generation space, \eg
$\lw1\lu \equiv \l1ijk H^c_iQ_ju^c_k\>$
with $i,j,k = 1,2,3$ generation indices,
and  in general they are not constrained by the
$\E$ Clebsch-Gordan relations.\upcite{witten-yuk}
As it stands the model is not phenomenologically
viable, since the simultaneous presence of $W_2$
and $W_3$
induces fast proton decay, and at the same time
the presence of $W_4$ would produce (too large) tree level
Dirac masses for all the neutral states.
Both these problems can be cured by imposing
on the superpotential \req{2.8} a discrete symmetry.
Such a symmetry must be generation dependent if
we want to leave open the possibility of having small neutrino masses
generated by loop effects.\upcite{branco}
\vglue 0.6cm
\adv{1}
\line{\elevenbf 3. Neutrino Masses in the Unconventional Schemes. \hfil}
\bigskip
\noindent
As it is clear from the second lines in
\req{2.4} and \req{2.5},
there is an ambiguity in assigning the known states to the {\bf 27}
representation, since under the SM gauge group
the ${\bf \bar 5}_{{\bf 10}}$ in
the {\bf 10} of $SO(10)$ has the same field content as the ${\bf
\bar 5}_{{\bf 16}}$ in the {\bf 16}. The same ambiguity is
also present for the two $\G$ singlets, namely
 ${\bf 1}_{{\bf 1}}$ and ${\bf 1}_{{\bf 16}}$.
Then, as a starting point for investigating $\E$ models with
UA, we will assume that what we call ``$\nu_\tau$"
is in fact the $N_3$ weak doublet neutral state belonging to the
${\bf \bar 5}_{{\bf 10}}$, while
\nue and \num are still assigned as usual to the
${\bf \bar 5}_{{\bf 16}}$.
We will henceforth use quotation marks to denote the known states with
their  conventional labels, since they might not correspond to the
entries in \req{2.4}-\req{2.6}. Labels not enclosed within quotation marks will
always refer to the fields listed in these equations.
Then, referring to the {\bf 10} and {\bf 16}
of $SO(10)$, our starting assumption for the assignments of the
three $SU(2)$ doublet light neutrinos reads:
$$
\eqalign{
&``\nu_\alpha" \in L_\alpha \in {\bf 16} \qquad\qquad \alpha=1,2 \cr
&``\nu_\tau"  \in H_3 \in {\bf 10}. \cr            }
\eq{3.1}
$$
\mn
In order to realise this scenario we first have to require
that the tree level masses for
$\nu_\alpha$ and $N_3$ vanish. This can be achieved by setting
in $W_4$
$$
\l{10}{\la i\ra}\alpha j (H^c_i L_\alpha \nu^c_j) = 0
\qquad {\rm and} \qquad
\l{11}{\la i\ra}3j       (H^c_i H_3      S^c_j) = 0.
\eq{3.2}
$$
\mn
For the sake of clarity we have enclosed inside
$\la$brackets$\ra$
the indices labeling the particular vacuum expectation values
which are relevant
for the actual discussion.
{}From the LEP measurement of the number of weak-doublet neutrinos we
know that all the remaining $SU(2)$ doublet neutral states
$N_\alpha$, $\nu_3$ and $N^c_i$ must be heavy ($\gtrsim 50\,$GeV).
This in turn implies that the following terms must be non-vanishing :
$$
\l{10}i3{\la\beta\ra}      (H^c_i L_3     \nu^c_\beta) \neq 0,
\qquad\qquad
\l{11}i\alpha{\la\beta\ra} (H^c_i H_\alpha  S^c_\beta) \neq 0.
\eq{3.3}
$$
\noindent
Now, in order to allow for radiatively generated Dirac masses, we need
massless R-handed neutrinos as well.
For the sake of simplicity, we will require
a {\elevenit minimum} number of light neutral
$SU(2)$ singlets.
In \req{3.3} we have already assumed that the couplings involving
$\nu^c_3$ and $S^c_3$ are forbidden, thus preventing
their fermionic component from acquiring a mass at tree level, so that
at the lowest order three $SU(2)$-doublet and two $SU(2)$-singlet
neutral states
are massless, namely $\nu_\alpha$ ($\alpha=1,2$),
$N_3$ and $\nu^c_3$, $S^c_3$.
Dirac masses for these states
can be induced by loops involving quarks, through the Yukawa couplings
appearing in $W_3$ in the superpotential \req{2.8}.
\vfill\eject
\null

\vskip 4.5truecm

\midinsert {
$$
\vbox{\hsize=10.3truecm
\baselineskip=10pt
\noindent Fig. 1:
{\tenrm
A typical diagram contributing to the neutrino Dirac masses
at the one-loop level.}
}
$$
\medskip
}
\endinsert
\interlinea\noindent
As is discussed in detail in Ref. [\cite{f}]
a set of one loop
diagrams analogous to the one depicted in Fig. 1 will generate
Dirac mass terms connecting the three L-handed neutrino with the
singlet $\nu^c_3$, while at this order $S^c_3$ remains decoupled.
However, additional diagrams are generated at the two loop level
through diagrams similar to the one depicted in Fig. 2, and they
induce additional mass terms for $S^c_3$ as well.

\vskip 4.5truecm

\midinsert {
$$
\vbox{\hsize= 10.3truecm
\baselineskip=10pt
\noindent Fig. 2:
{\tenrm
A two-loop diagram giving rise to $\nu_\alpha$-$S^c_3$ entries
in the neutrino Dirac mass matrix.}
}
$$
}
\endinsert
\interlinea

The final form of the mass matrix reads
$$
(\nu_1\>\nu_2\>N_3) \, \cdot {\cal M} \cdot
\pmatrix{0 \cr S^c_3 \cr \nu^c_3 \cr},
\hskip 2.3 truecm
{\cal M} =
\pmatrix{ 0&b_1&a_1 \cr 0&b_2&a_2 \cr 0&0&a_3 \cr}
\eq{3.4}
$$
and is unique for the minimal scheme we have chosen\upcite{f}.
The entries $a_1$, $a_2$ and $a_3$ are generated at one loop,
and with a reasonable choice of the parameters entering the
diagram in Fig.~1 they
can be estimated to fall naturally in the range $0.1-10$ eV.
The entries $b_1$ and $b_2$ are generated at two loop,
thus acquiring a typical suppression factor of order 10$^{-3}$
with respect to the one loop masses.
Once the mass matrix is diagonalized, a very interesting pattern
of mixings and
masses for the eigenstates $n_1$, $n_2$ and $n_3$ results.\upcite{f}
We end up with a massive
neutrino $n_3$, mainly ``$\nu_\tau$", with a mass
that can naturally fall in the range
$\sim 0.1-10\,$eV. The lower value is interesting for
``$\nu_\mu$"-``$\nu_\tau$" atmospheric neutrino oscillation.
On the other hand, due to the fact that $n_3$
is cosmologically
stable,\upcite{f}  if its mass were close to the upper value,
it could provide an interesting
candidate for the hot component of the dark matter (DM).
A second neutrino $n_2$, mainly ``$\nu_\mu$",
acquires a much smaller mass $(\sim 10^{-3})$
at the two
loop level, and can be relevant for matter enhanced ``\nue''-``\num''
oscillations in the sun.\upcite{MSW} Finally,
due to the absence in our minimal scheme of
a third helicity partner for the L-handed neutrinos,
$n_1$ remains massless.
We stress that such a hierarchy of masses (and
a corresponding hierarchy of mixings\upcite{f})
arises naturally as a direct consequence of the UA, and reflects the
fact that being the neutrinos embedded in a different way in the gauge
group, for one species a Dirac mass is generated at one-loop, while
for the other two species the corresponding diagrams are forbidden,
and only at the two loop level can a mass arise.
\vglue 0.6cm
\adv{1}
\line{\elevenbf 4. Conclusions. \hfil}
\medskip
\noindent
In conclusion we have described the possibility of constructing
consistent models in which the known neutrinos of the three different
generations do not have the same gauge interactions under possible
additional $U(1)^\pr$ factors. We have carried out our analysis in the
frame of the superstring--inspired $\E$ models, taking as a guideline
the requirement of having interesting
neutrino phenomenology with naturally small radiatively generated
Dirac masses. Models based on this scheme are
indeed viable and can be realised by imposing a family-non-blind
discrete symmetry on the superpotential.\upcite{f}
We have briefly described a minimal model, in which only two
additional light $SU(2)$ singlet neutrinos are present, thus leaving
one doublet neutrino massless.
Clearly other models based on the same scenario but
with a more rich structure in the neutrino sector can
also be constructed.
We have stressed that values of
the neutrino masses in ranges interesting for explaining
the solar and the atmospheric neutrino anomalies,
or possibly for providing a hot DM component,
can be obtained with a natural choice of the parameters.
A hierarchy of masses and mixing is naturally generated
as a consequence of the choice of UA assignments.
We stress that, as it has been recently discussed,
signals of models predicting UA could be detected at future $e^+e^-$
colliders by measuring ratios of cross sections for lepton
productions.\upcite{fphen} Also the correct pattern of embedding
of the leptons into the gauge group could be identified through
a large set of experimental observables
including asymmetry measurements.\upcite{e6asy}
\vglue 0.6cm
\vfill\eject
\adv{1}
\line{\elevenbf 5. Acknowledgements \hfil}
\vglue 0.4cm
\noindent
It is a pleasure to thank J.W.F Valle, A. Perez, D. Tommasini
and F. Campos for inviting me at the
{\elevenit International School on Cosmological Dark Matter}
and for making my stay in Valencia very pleasant.
\vglue 0.6cm
\line{\elevenbf 6. References \hfil}
\vglue 0.4cm

\biblitem{see-saw}
M. Gell-Mann, P. Ramond, and R. Slansky, in {\elevenit
Supergravity}, F. van Nieuwenhuizen and D. Freedman eds., (North
Holland, Amsterdam, 1979) p.~315; \hbup
T. Yanagida, {\elevenit Proc. of the
Workshop on Unified Theory and Baryon Number of the Universe}, KEK,
Japan, 1979.  \par

\biblitem{rizzo-e6}
For a review see J.L. Hewett and T.G. Rizzo, \prep{183} 195 (1989).
\par

\biblitem{ellis-e6}
B.A. Campbell \ea, \ijmpa{2} 831 (1987). \par

\biblitem{MNS}
A. Masiero \ea, 
\prl{57} 663 (1986).  \par

\biblitem{branco}
G.C. Branco and C.Q. Geng, \prl{58} 969 (1986).\par

\biblitem{f}
E. Nardi, \prd{48} 3277 (1993). \par

\biblitem{fphen}
E. Nardi, Report UM-TH-93-19, to appear on Phys. Rev. {\elevenbf D}. \par

\biblitem{e6asy}
E. Nardi and T.G. Rizzo, Report SLAC-PUB--6422 and  UM-TH-93-29.
\par

\biblitem{e6-super}
D. Gross, J. Harvey, E. Martinec and R. Rhom, \prl{54} 502 (1985);
\npb{256} 253 (1985); \ib {\elevenbf B 267} (1986) 75; \hbup
P. Candelas, G. Horowitz, A. Strominger and E. Witten, \npb{258}
46 (1985); E. Witten, \plb{149} 351 (1984). \par

\biblitem{witten-yuk}
E. Witten, \npb{258} 75 (1985). \par

\biblitem{MSW}
L.~Wolfenstein, \prd{17} 2369 {1978}; {\elevenbf D20}, 2634 (1979); \hbup
S.~P.~Mikheyev and A.~Yu.~Smirnov,
{\elevenit Yad.  Fiz.} {\elevenbf 42}, 1441 (1985); \ncim{{9C}} 17 (1986).\par

\biblitem{weinberg-fc}
S. Glashow and S. Weinberg, \prd{15} 1958 (1977).\par

\interlinea
\insertbibliografia

\vfill\eject\bye